# Generic System of Exact Flavor-Electroweak Quantities via New Dynamical Parameter

E. M. Lipmanov
40 Wallingford Road # 272, Brighton MA 02135, USA

Abstract

Unification-type idea of a basic connection between particle mass and charge quantities is in the spirit of string theory. On the level of phenomenology, this idea means raison d'etre for particle flavor since it requires more than one mass copy i.e. transition from individual particle mass to the concept of mass-matrix. In this paper that idea is exemplified by a generic system of accurate empirical relations for dimensionless flavor-electroweak quantities (fine structure constant, muon-electron and tauon-electron mass ratios and quark CKM and neutrino PMNS mixing angles) build in terms of one small universal $\varepsilon$-parameter as mediator of mass-matrix and charge quantities. The used in this study empirically suggested math paradigm consists of repeated exponentiations of $\varepsilon$-powers. Sharp accuracy boost of flavor relations from replacement of $\varepsilon$-power terms by exponential f-terms is observed. The two widely discussed in the literature empirical flavor regularities, quark-lepton complementarity (QLC) and Koide charged lepton mass formula, are essential parts of the generic system. Solar neutrino mixing angle is predicted $\theta_{sol} \cong 34.05°$ by extension of QLC. Charged lepton mass ratios satisfy Koide relation with high accuracy $\sim 10^{-9}$. The Appendix contains comments on dual objective-anthropic nature of physical reality.

# 1. Introduction.

There are no exact predictions for the flavor-electroweak mass and mixing parameters and gauge constants in the Standard Model. And no serious indications to date that the accurate empirical flavor regularities including particle mass ratios and mixing angles can be derived from discussed symmetries at high energy scale without emergence of a new basic dimensionless physical constant[1]. If something is missing, it may be that it is the relations between the particle flavor quantities (mass-matrices) and universal charge.

In this paper we consider empirical indications on flavor-electroweak (FEW) connections in terms of one small universal empirical FEW parameter [3],

$$\varepsilon = \exp(-5/2). \qquad (1)$$

The unification of dimensionless mass-matrix quantities and universal electric charge is exemplified below in terms of this dimensionless parameter. Such unification becomes possible only if there is more than one particle mass copy and so it means raison d'etre in flavor phenomenology. A large accurate generic system of dimensionless particle flavor and EW quantities – fine structure constant, charged lepton mass ratios and quark and neutrino mixing angles – is studied by primary analyses of relevant experimental data.

The concept of flavor benchmark pattern [4] (zero $\varepsilon$-approximation) is used as the point of departure in the discussed flavor phenomenology.

---

[1] The Cabibbo angle is usually mentioned as such type of constant, e.g. [1, 2].

A special empirical feature of the present study is that the approximately fitting data small ε-power terms (perturbative deviations from benchmark values) of the flavor quantities are replaced by one universal exponential function $f(x) = x e^x$ with the argument x in terms of those ε-powers. These replacements boost the accuracy of considered flavor quantities compared to data values.

Accurate FEW relations build from ingredients related to the ε-parameter (1), small integers and quantities π and √2 are discussed in Sections 1-9. Conclusions are in Sec.10. In the Appendix a toy model of anthropic selection is considered (it may be skipped).

## 2. Fine structure constant, $α ≡ α(q^2 = 0)$

At benchmark α = 0; at leading ε-approximation $α = ε^2$ to within ~8%. An accurate equation (comp. [3]) for the low energy fine structure constant at next to leading ε-approximation is given by

$$(\exp α / α)^{(\exp 2α)} - f(-α/π) = 1/ε^2 \quad (2)$$

with function f(x) defined as

$$f(x) = x \exp(x). \quad (3)$$

By PDG-2010 data [6] $α_{exp} = 1/137.035999679(94)$ the relation (2) is accurate to within ~$3×10^{-9}$. Numerical solution of the equation (2) with ε as source yields

$$α = 1/137.0360001451. \quad (4)$$

This approximate solution of Eq.(2) is to within ~$2×10^{-13}$.

Solution (4) is in agreement with PDG-2010 α-value with high accuracy ~**5** S.D.

A small modification ($\alpha \to \varepsilon^2$ in the exponent) of the exponential function $f(-\alpha/\pi)$ on the left side of Eq.(2) leads to a more accurate (against data [6]), equation

$$(\exp \alpha /\alpha)^{(\exp 2\alpha)} + (\alpha/\pi)\exp(-\varepsilon^2/\pi) = 1/\varepsilon^2. \qquad (5)$$

Its accuracy is enhanced to $\sim 4 \times 10^{-10}$. The numerical solution of equation (5) is

$$\alpha = 1/137.0359997372. \qquad (6)$$

This is an approximate solution of Eq.(5) accurate to within $1 \times 10^{-13}$.

The solution (6) is in high accurate agreement with PDG-2010 $\alpha$-value – to within ~**0.6** S.D.

It should be noticed that the recently published new experimental value for the fine structure constant [12] $\alpha^{-1}_{exp} = 137.035999037(91)$, which agrees with previous one [13] $\alpha^{-1}_{exp} = 137.035999084(51)$, can be accurately described[2] by the exact Eq.(21'') on p.10 of ref. [3]:

$$(\exp \alpha /\alpha)^{(\exp 2\alpha)} - (\alpha/\pi)\exp(-\alpha/4) = 1/\varepsilon^2. \qquad (7)$$

The solution of Eq.(7) is given by

$$\alpha = 1/137.035999003. \qquad (8)$$

It is accurate to within ~0.4 S.D. by comparison with data of ref. [12], and ~1.6 S.D. by comparison with ref. [13].

Finally, the formal quantitative structure of the accurate equation (5), or (7), for the $\alpha$–constant may be approximately presented in the form

$$(1/\alpha)[\alpha (\exp \alpha /\alpha)^{(\exp 2\alpha)} + (\alpha^2/\pi)\exp(-\varepsilon^2/\pi, \text{ or } -\alpha/4)] \cong$$
$$(1/\alpha)[1.08 + 0.38\, \varepsilon^4] \cong (1/\varepsilon^2), \qquad (9)$$

or by inverse order

$$(1/\alpha) \cong [0.92 - 0.32\, \varepsilon^4](1/\varepsilon^2). \qquad (10)$$

---

[2] The small term in equation (7) can be rewritten in terms of function (3): $(\alpha/\pi)\exp(-\alpha/4) = -(4/\pi)f(-\alpha/4)$.

The quantitative ε-structure of the inverse fine structure constant (10) appears a useful guide in search for accurate relations for the charged lepton mass ratios below.

### 3. Quark mixing angles

The start is with *zero* ε-approximation in the form of a unit quark benchmark [4] mixing matrix:

$$\sin^2(2\theta_c) = 0, \quad \sin^2(2\theta_{23}) = 0, \quad \sin^2(2\theta_{13}) = 0, \quad \theta_c = \theta_{12}. \quad (11)$$

Fitting the data leading small finite ε-power deviation from benchmark mixing (11) is given by

$$\sin^2(2\theta_c) \cong 2\varepsilon, \quad \sin^2(2\theta_{23}) \cong \varepsilon^2, \quad \sin^2(2\theta_{13}) \cong \varepsilon^4,$$

$$\theta_c \cong 11.9°, \quad \theta_{23} \cong 2.4°, \quad \theta_{13} \cong 0.19°. \quad (12)$$

Finally, a pattern of three quark CKM mixing angles *at* next to leading ε-approximation is obtained by replacing the right-hand ε-power terms in the leading approximation (12) by f-functions (3) with that terms as arguments [5],

$$\sin^2(2\theta_c) \cong f(2\varepsilon) = (2\varepsilon)\exp(2\varepsilon), \quad \theta_c \cong 13.047°, \quad (13)$$

$$\sin^2(2\theta_{23}) \cong f(\varepsilon^2) = (\varepsilon^2)\exp(\varepsilon^2), \quad \theta_{23} \cong 2.362°, \quad (14)$$

$$\sin^2(2\theta_{13}) \cong f(\varepsilon^4) = (\varepsilon^4)\exp(\varepsilon^4), \quad \theta_{13} \cong 0.193°. \quad (15)$$

Solutions (13)-(15) for quark mixing angles are accurate to within small 1 S.D. of the quark CKM world fit mixing matrix [6] – an accuracy boost by comparison with (12) especially for the Cabibbo angle.

The high accuracy and regularity of the pattern (13)-(15) for quark CKM mixing angles are remarkable.

### 4. Neutrino mixing angles

The start is with the benchmark flavor pattern [4] in the form of bimaximal neutrino mixing matrix,

$$\cos^2(2\theta_{12}) = 0, \quad \cos^2(2\theta_{23}) = 0, \quad \sin^2(2\theta_{13}) = 0, \quad (16)$$

$$\begin{pmatrix} 1/\sqrt{2} & 1/\sqrt{2} & 0 \\ -1/2 & 1/2 & 1/\sqrt{2} \\ 1/2 & -1/2 & 1/\sqrt{2} \end{pmatrix} \nu. \quad (16')$$

It is a suggestive zero $\varepsilon$-approximation of neutrino mixing. Together with the zero approximation unit quark mixing matrix (11) it presents an extreme form of quark-lepton complementarity [7]. Fitting the data leading (perturbative) $\varepsilon$-approximation for the empirically large neutrino mixing angles is given by

$$\cos^2(2\theta_{12}) = 2\varepsilon, \quad \cos^2(2\theta_{23}) = \varepsilon^2, \quad \sin^2(2\theta_{13}) \cong 2\varepsilon^2,$$

$$\theta_{12} \cong 33°, \quad \theta_{23} \cong 42.6°, \quad \theta_{13} \cong 3.3°. \quad (17)$$

The first two relations for neutrino angles $\theta_{12}$ and $\theta_{23}$ are from the leading $\varepsilon$-approximations for quark $\theta_c$ and $\theta_{23}$ ones in Sec.2 via exact QLC relations; in contrast, the neutrino mixing angle $\theta^{\nu}_{13}$ is related [4] to the quark one $\theta^{q}_{13}$ by the quadratic hierarchy rule $\sin^2(2\theta^{\nu}_{13}) \cong 2\sin(2\theta^{q}_{13})$.

Using the accurate presentation of the quark CKM mixing angles (13)-(15), 'universal' deviation from exact QLC at next to leading $\varepsilon$-approximation for solar and atmospheric mixing angles is suggested [5],

$$\cos^2(2\theta_{sol}) = f(2\varepsilon)e^{-4\varepsilon}, \quad \theta_{sol} \cong 34.04°, \quad (18)$$

$$\cos^2(2\theta_{atm}) = f(\varepsilon^2)e^{-4\varepsilon}, \quad \theta_{atm} \cong 43°. \quad (19)$$

These relations connect the two largest mixing angles in neutrino and quark mixing matrices in fair agreement with data analyses [8, 9].

Another interesting deviation from exact geometric QLC is 'combined QLC' considered in [11]. Unlike universal deviation from QLC it is fully expressed in terms of the f-function in analogy with the relations for the quark mixing angles. Instead of (18) and (19), the solar and atmospheric mixing angles are determined now by relations

$$\cos^2(2\theta_{sol}) = |f(-2\varepsilon)|, \quad \theta_{sol} \cong 34.04°, \quad (20)$$

$$\cos^2(2\theta_{atm}) = f(\varepsilon^2), \quad \theta_{atm} \cong 42.64°. \quad (21)$$

The only deviation from exact QLC in (20),(21) is the opposite sign of $\varepsilon$-parameter in comparison with that for quark mixing angles.

It follows from (20),(21) that i) the finite deviation from QLC of the solar mixing angle is the only QLC-violation and its predicted magnitude is not change in comparison with (18)), ii) the atmospheric mixing angle obeys exact QLC and its value is slightly degreased from 43° in (19) to $\theta_{atm} \cong 42.64°$ in (21); the deviation of the atmospheric mixing angle from maximal mixing is $\delta_{23} = (45-\theta_{23}) \cong 2.36°$ instead of $\delta_{23} \cong 2°$ in (19).

The neutrino solar and atmospheric mixing angles (18)-(19), or (20)-(21), appear in especially good agreement with the best fit values from the analysis of oscillation data by Fogly et al [8] if taken for granted. They are testable and definitely falsifiable by coming accurate neutrino oscillation data.

## 5. Muon-electron DMD-quantity

At zero $\varepsilon$-approximation the mu/e mass ratio is given by benchmark flavor pattern [4] as infinitely large. At finite $\varepsilon$-parameter the large deviation from mass-degeneracy (DMD) mu/e-quantity should be $(m_\mu/m_e - 1) = \sqrt{2}/\varepsilon^2$, accuracy ~2%.

At next to leading perturbative $\varepsilon$-approximation it is built in terms of the $\varepsilon$-parameter and by analogy with the accurate quantitative structure (10) for the fine structure constant $\alpha$ in Sec.2 as

$$(m_\mu/m_e - 1) \cong (\sqrt{2}/\varepsilon^2)[1 - 3\varepsilon^2 - (\sqrt{2}-1)\varepsilon^4/\sqrt{2}] \cong 205.687854. \qquad (22)$$

It is accurate to within $\sim 4 \times 10^{-4}$.

By replacement of all small $\varepsilon$-powers in (22) with the small f-function terms, an accurate form of the mu/e-DMD-quantity follows

$$(m_\mu/m_e - 1) \cong (\sqrt{2}/\varepsilon^2)\{1 + f(-3\varepsilon^2) - f[(\sqrt{2}-1)\varepsilon^4/\sqrt{2}]\} \cong$$
$$(\sqrt{2}/\varepsilon^2)\{\exp f(-3\varepsilon^2) - f[(\sqrt{2}-1)\varepsilon^4/\sqrt{2}]\}. \qquad (23)$$

The function $\exp f(x)$ is a twice exponential one, $\exp f(x) = \exp(x)^{\exp(x)}$; at small x its value is close to unity.

Relation (23) predicts muon-electron mass ratio and absolute muon mass through the $\varepsilon$-parameter and high accurate data value of the electron mass $(m_e)_{exp} = 0.510998910 \pm 1.3 \times 10^{-8}$ MeV [6]:

$$m_\mu/m_e \cong 206.768280353, \qquad (24)$$

$$m_\mu \cong 105.65836588 \pm 2.7 \times 10^{-6} \text{ MeV}. \qquad (25)$$

The value (25) is in high accurate agreement with the best fit value of PDG mu-mass [6], $(m_\mu)_{exp} = 105.658367 \pm 4 \times 10^{-6}$ MeV, to within $\sim 0.7$ S.D. ($1 \times 10^{-8}$; the boost in accuracy against the $\varepsilon$-power approximation (22) is about four orders of magnitude).

The mu/e DMD-relation (23) may be represented in approximate form similar to the quantitative structure of the relation (10) for $\alpha$:

$$(m_\mu/m_e - 1) \cong (\sqrt{2}/\varepsilon^2)[0.98 - 0.3\varepsilon^4]. \qquad (26)$$

Two interesting inferences should be emphasized.
i) There is a close quantitative analogy between the structure of the next to leading small $\varepsilon$-corrections in the square bracket of muon-electron DMD value (26) and the one (10) for the inverse fine structure constant: $\sim(1 - 0.31\varepsilon^4)$ and $\sim(1 - 0.35\varepsilon^4)$ respectively, ii) this analogy appears only if the muon-electron DMD-quantity is the primary physical one; in contrast, if the mu/e mass ratio is used as primary, it disappears.

In conclusion, it is interesting to note that the appearance of the f-function terms (f-terms) in the high accurate mu/e-relation (23) follows the pattern observed in the relevant cases of quark and neutrino mixing angles. Starting with zero approximation benchmark values for flavor quantities, approximate small expressions in $\varepsilon$-powers are built as first step. At second step, the $\varepsilon$-power terms are replaced by f-function terms having those $\varepsilon$-powers as arguments, i.e. compare (22) and (23). These replacements lead always to large gains in accuracy.

## 6. Tauon-electron DMD-quantity

At zero $\varepsilon$-approximation it is infinitely large (benchmark pattern [4]). At finite $\varepsilon$-parameter it is large $(m_\tau/m_e - 1)$ = **$2/\varepsilon^3$**. At next to leading $\varepsilon$-approximation it can be presented in f-terms quite similar to the accurate empirical relation (23) in mu/e case,

$$(m_\tau/m_e - 1) \cong (2/\varepsilon^3) [\exp f(-\varepsilon/2) + f(\varepsilon^4/5)]. \qquad (27)$$

It should be underlined the economy-feature[3] of the expressions in f-terms of charged lepton mass ratios (23) and (27). They do not include any additional numerical coefficients accompanying the f-terms. This feature is shared by relations (13)-(15) and (20)-(21) for quark and neutrino mixing angles.

The τ/e mass ratio and prediction for the absolute τ-lepton mass are

$$m_\tau/m_e \cong 3477.441573159, \qquad (28)$$

$$m_\tau \cong 1776.968853 \text{ MeV}. \qquad (29)$$

---

[3] Compare with the corresponding relations in the previous versions of this arXiv-publication.

The tau/e DMD $\varepsilon$-structure that is similar to the approximate quantitative structure of the relation (10) for $1/\alpha$ is given by

$$(m_\tau/m_e - 1) \cong (2/\varepsilon^3)[\,0.96 + 0.2\,\varepsilon^4\,]\,. \qquad (30)$$

The three $\varepsilon$-relations obtained above for the inverse fine structure constant $1/\alpha$ (5),(7), $m_\mu/m_e$ mass ratio (23) and $m_\tau/m_e$ one (27) have similar quantitative $\varepsilon$-structures - $(1 - 0.35\,\varepsilon^4)$, $(1 - 0.31\,\varepsilon^4)$, $(1 + 0.21\,\varepsilon^4)$ respectively. Probably, they may point to a common dynamical origin of the small $(\varepsilon^4)$-terms in these estimations.

### 7. Quadratic hierarchy regularity

An interesting inference from the obtained pair of accurate charged lepton mass ratios (24) and (28) is that they satisfy the quadratic hierarchy flavor rule applied to the two arithmetic mean values of charged lepton mass,

$$<m> = (m_e + m_\mu + m_\tau)/3 = m_e(1 + m_\mu/m_e + m_\tau/m_e)/3, \qquad (31)$$

and square-root-mass,

$$<\sqrt{m}> = (\sqrt{m_e} + \sqrt{m_\mu} + \sqrt{m_\tau})/3 = \sqrt{m_e}(1 + \sqrt{m_\mu/m_e} + \sqrt{m_\tau/m_e})/3, \qquad (32)$$

averaged over the three lepton flavor states. That quadratic hierarchy rule is given here by relation

$$<m> = 2(<\sqrt{m}>)^2. \qquad (33)$$

As a particular quantitative result, this quadratic hierarchy rule is satisfied by the charged lepton mass ratios (24) and (28) with high accuracy,

$$[<m> - 2(<\sqrt{m}>)^2]/<m> \cong 1\times10^{-9}, \qquad (34)$$

which is independent of the electron mass $m_e$ in the average-relations (31) and (32).

Notice that this interesting result is mainly due to conditions: 1) right choice of the leading perturbative $\varepsilon$-

approximations for lepton DMD-quantities based on the benchmark flavor pattern, 2) replacement of ε-powers by exponential f-terms and 3) similar quantitative structures $\sim(1\pm 10^{-5})$ of the small next to leading ε-approximations for the three basic flavor-electroweak quantities: $(1/\alpha)$, $(m_\mu/m_e - 1)$ and $(m_\tau/m_e - 1)$.

## 8. Relation to Koide formula [10]

The τ-lepton mass magnitude (29) is close to the value predicted earlier by Koide from the suggestion that the Koide charged lepton pole mass formula,

$$(m_e + m_\mu + m_\tau) = (2/3)(\sqrt{m_e} + \sqrt{m_\mu} + \sqrt{m_\tau})^2, \quad (35)$$

$$(1 + m_\mu/m_e + m_\tau/m_e) = (2/3)[1 + \sqrt{(m_\mu/m_e)} + \sqrt{(m_\tau/m_e)}]^2, \quad (35')$$

is an exact one, and input of the PDG best-fit value of muon-electron mass ratio and absolute mass of the electron into equation (35). And it should be so indeed since the quadratic hierarchy rule (33) in the realistic case of three flavor generations coincides with the Koide relation (35).

The advantage of the start with charged lepton DMD-quantities and quadratic hierarchy paradigm is that it answers the three main questions: a) why is the Koide relation symmetric over the three charged lepton masses, b) why does it agree with the data only in the quadratic form, but not in other power-forms, and c) why the integer 3 in the denominator of (35). The Koide prediction for the τ-lepton mass $m_\tau \cong 1776.97$ MeV is still not excluded by experimental data.

## 9. The Weinberg weak mixing angle

Consider another basic SM dimensionless quantity [6],

$$[\sin^2\theta_W(M_Z)(MS)]_{exp} = 0.23116(13). \quad (36)$$

At leading ε-approximation it is given by $\sin^2\theta_W(M_Z) = f(1/5) \cong 0.24428$. At next to leading ε-approximation, it may be represented[4] as

$$\sin^2\theta_W(M_Z) = f(1/5) + (2/5)f[f(-5)] \cong 0.23125. \quad (37)$$

The result (37) agrees with the experimental one (36) to within 0.7 S.D.

## 10. Some inferences and conclusions

**1**. The main principal result of the present research is an exemplified answer to one of the fundamental problems of particle physics – why is particle flavor needed at all. The need of particle flavor (degree of freedom) can be motivated by postulating an attractive new unification condition - connection between particle dimensionless mass-matrix quantities and the universal charge.

With one particle flavor there is no relation between particle mass and charge. In contrast, with extra flavor mass copies, as shown above, there are empirical connections between dimensionless flavor quantities and the universal particle charge. Accurate connections between particle mass ratios and mixing angles and dimensionless made electric charge are found from analysis of empirical data in terms of the new universal ε-parameter.

**2**. That the fine structure constant and weak mixing angle are determined by the ε-parameter, (2) and (37), means that all the dynamical gauge constants (charges) of the EW theory are expressed through the ε-constant. Therefore, it is quite natural to suggest, and especially in view of the discussed above accurate generic system of flavor-EW

---

[4] In terms of the ε-parameter:
$$\sin^2\theta_W(M_Z) = f(1/\log\varepsilon^{-2}) - (1/\log\varepsilon)f[f(\log\varepsilon^2)]. \quad (37')$$

relations, that the unknown dynamical constant of new flavor physics should be also closely related to that ε-constant.

**3**. Two high accurate relations for the fine structure constant are discussed in Sec.2 that agree with the 1) PDG-2010 experimental data with accuracy to within ~**0.6** S.D. and 2) special data in refs. [12] and [13] with accuracy to within ~0.4 S.D. and ~1.6 S.D. respectively.

**4**. Considered above high accurate low energy regularities between dimensionless flavor quantities and fine structure constant are without radiative corrections or, on the contrary, include already the main radiative corrections. This observation may be related to a new kind of 'strange' physical conditions – anthropic selections commented in the Appendics.

**5**. In three cases $(1/\alpha)$, $(m_\mu/m_e - 1)$ and $(m_\tau/m_e - 1)$ the next to leading exponential factors are divided in two different parts: 1) twice exponential parts that are close to unity and 2) exponential parts that are much smaller ones with nearly equal relative magnitudes in all three cases estimated as $\sim (\varepsilon^2)^2 \sim 10^{-5}$. Probably, those two parts are from very different physics sources.

**6**. With benchmark flavor pattern as point of departure, a pattern of replacements of the small ε-power corrections (perturbative-type) to benchmark flavor quantities by small exponential f-terms is considered for a large system of particle flavor quantities. The used in this study empirically suggested math paradigm consists of repeated exponentiations of ε-powers with coefficients from small integers. Interesting empirical effect of those replacements is a remarkably large accuracy enhancement (accuracy boost) of all considered flavor quantities. If this empirical effect gains further support, it may point to new flavor physics paradigm hidden behind the extra flavor generations (as cause of particle mass matrices).

**7**. A suggested inference by accurate lepton mass ratios (23) and (27) is that they satisfy the Koide charged lepton mass relation with high accuracy $\sim 1\times 10^{-9}$. The predicted accurate value of the τ-lepton mass is close to the one predicted earlier by Koide $m_\tau \cong 1776.97$ MeV.

## **Appendix**
### **A. A toy model of anthropic selection**

In contrast to objective physical realities, there are 'conceptual' realities as not directly observable notions that are successfully used in theory as backgrounds for physical quantities. In general, anthropic selection may have meaning in a symmetric manifold of conceptual universes containing the universe, or universes, suitable for life. Emergence of life and consciousness in one particular universe is a spontaneous symmetry violation of the manifold of conceptual universes. It singles out the one particular Universe U as dual objective-anthropic physical reality for 'our' community of observers[5].

The ultimate free parameters in fundamental physics are appropriate places where the anthropic meaning of our Universe may be revealed. As an illustration, consider a toy model of anthropic selection in particular relation to the universal dimensionless $\varepsilon$-constant and discussed above system of accurate flavor-electroweak regularities.

Let us define conceptual $\varepsilon_n$-parameters in a manifold of conceptual universes $U_n$ numerated by not-negative integer numbers 0, 1, 2, 3... n ... These parameters are given by

$$\varepsilon_n = \exp(-n/2), \quad 0 < \varepsilon_n < 1. \tag{A.1}$$

---

[5] By the definition, there is only one objective-anthropic Universe. Even if there are several identical fitting to life conceptual universes, but with no connections between them, 'our' Universe would be the only physical reality. If an empirical discovery appears in favor of an observable connection between our Universe and some conceptual one, the latter will become a part of our Universe.

The system of considered above flavor-electroweak quantities is described by small-parameter perturbations of the exact benchmark flavor pattern [4] associated with the extreme conceptual universe $U_\infty$. The final results in Secs.2-9 are at $\varepsilon = \varepsilon_n|_{n=5}$, Eq.(1), in our Universe $U = U_5$.

In this model, anthropic selection of our Universe from the manifold of conceptual universes means spontaneous violation of the arithmetic symmetry of integer numbers by singling out $n = 5$ for $\varepsilon_n$-parameter through emergence of life in the Universe $U_5$. Finite integers that are close to integer 5 are also singled out[6][7].

On the background of benchmark flavor pattern, small value of the fine structure constant is related to large charged lepton mass ratios by generic quantitative connections of these quantities to the $\varepsilon$-constant, which leads to relations $m_\tau/m_\mu \cong \sqrt{(2/\alpha)}$, $m_\mu/m_e \cong \sqrt{2}/\alpha$ at leading approximation.

## B. On dual objective-anthropic nature of physical reality

What is objective reality has a straightforward answer only in religion; a consistent definition is given by the grate Irish philosopher George Berkeley (18-th century): objective reality is that observed by god. The term 'god' is not appropriate in physics. Physics is based on the principle that all knowledge is from processed observations. So, the processing observations intelligent community endowed with large memory stored in minds, books and computers and especially in the language of math may actually play Berkley's god-observer in the appropriate definition of objective physical reality.

Objective physical reality without anthropic ingredients is something not observable, which cannot enter physics. Whatever meaning one may want to attach to the notion of "existence of never observable (directly or indirectly[8]) objective reality", it cannot have other relation to physics than just convenience. To the point, this consideration was at the start of Einstein's relativity theory.

---

[6] In simple words, the small integers surrounding integer 5 are singled out in basic physics because of the anthropic nature of our universe.
[7] The suggested by empirical data math paradigm used in the present study of dimensionless flavor-electroweak quantities - after small perturbation of the benchmark flavor pattern as the point of departure - is based on mentioned *small integers and repeated exponentiations*.
[8] Quarks are considered objective physical reality since quark theory has many predictions confirmed by experimental data. If string theory had specific confirmed by data predictions, strings would be objective physical reality analogous to quarks.

Some observed low energy accurate physical regularities are probably nature's fundamental anthropic messages in physics. Accurate low energy fine structure constant and elementary particle pole mass ratios (free parameters in the Standard Model) as known are among the necessary conditions for life.

*A concept of dual objective-anthropic nature of physical reality fits well the facts.*

Objective aspect of physical reality is that it exists independent of the minds of observers. It is the regular presupposition of physics; moreover, the idea that objective reality *exists* independent of the mind is the most general and successful one in all science.

Anthropic aspect of physical reality includes both 1) certain objective conditions of life development do exist. They lead to the emergence of consciousness that spontaneously singles out the physical Universe from a manifold of conceptual universes[9], and
2) physical reality as perceived by observers is the source and contents of physics. By processing observations the physical community is still developing the *idea* of a Universe of objective physical realities, which existence is continued to the past with a likely start at the big bang.

---

[9] It makes that particular universe 'our Universe' no matter how terribly unlikely it may look in the manifold of conceptual universes.